\documentclass[preprintnumbers,aps,amsmath,amssymb,pra,twocolumn,showpacs,superscriptaddress]{revtex4-1}

\usepackage{enumerate}
\usepackage{amsmath}
\usepackage{amssymb}
\usepackage{graphicx}
\usepackage{color}
\usepackage{euscript}
\usepackage{float}

\newtheorem{Lemma}{Lemma}

\newcommand{\lv}{\left \vert}
\newcommand{\rv}{\right \vert}
\newcommand{\la}{\left \langle}
\newcommand{\ra}{\right \rangle}
\newcommand{\ket}[1]{\lv #1 \ra}

\newcommand{\ketbra}[2]{\lv #1 \ra \la #2 \rv}
\newcommand{\average}[1]{\la #1 \ra}

\newcommand{\Raverage}[1]{\la #1 \ra_{\mathrm{rand}}}
\newcommand{\averageInf}[1]{\la #1 \ra_{T;\infty}}
\newcommand{\Rsigma}{\sigma_{\mathrm{rand}}}
\newcommand{\EMW}{E_{\mathrm{MW}}}
\newcommand{\tr}{\mathrm{Tr}}

\begin{document}

\title{Simulating typical entanglement with many-body Hamiltonian dynamics}
\author{Yoshifumi Nakata}
\affiliation{Department of Physics, Graduate School of Science, University of Tokyo, Tokyo 113-0033, Japan}

\author{Mio Murao}
\affiliation{Department of Physics, Graduate School of Science, University of Tokyo, Tokyo 113-0033, Japan}
\affiliation{Institute for Nano Quantum Information Electronics, University of Tokyo, Tokyo 153-8505, Japan}

 \begin{abstract}
We study the time evolution of the amount of entanglement generated by one
dimensional spin-1/2 Ising-type Hamiltonians composed of
many-body interactions.
We investigate sets of states randomly selected during the time evolution generated by several types of time-independent Hamiltonians
by analyzing the distributions of the amount of entanglement of the sets.
We compare such entanglement distributions with that of typical entanglement, entanglement of a set of states randomly selected from a Hilbert space with respect to the unitarily invariant measure.
We show that the entanglement distribution obtained by a time-independent Hamiltonian can
simulate the average and standard deviation of the typical entanglement, if the Hamiltonian contains suitable many-body interactions.  We also show that the time required to achieve such a distribution is polynomial in the system size for certain types of Hamiltonians.
\end{abstract}

\date{\today}


 \maketitle
 \section{Introduction}
Random states are a set of pure states uniformly distributed in a Hilbert space with
respect to the unitarily invariant measure. In quantum information, random states are used in several quantum protocols, for example,
remote preparation of quantum states \cite{BHL2005},
quantum data hiding \cite{TDL2001,DLT2002} and quantum one-time pads \cite{HLSW2004}.
In physics, the properties of random states are studied in relation to a derivation of
a canonical distribution in quantum statistical mechanics \cite{GLTZ2006, PSW2006,LPSW2009}
and information leakage by the evaporation of black holes \cite{HP2007}.
One way to characterize a set of states is by analyzing
the distribution of the amount of entanglement of the states, often called the
{\it entanglement distribution}.
Recently, it has been shown that the entanglement distribution of random states
concentrates just below the maximum in several measures of entanglement \cite{HLW2006}.
Such characteristic appearing in random states is referred to as {\it typical entanglement}.

Generation of exact random states requires exponential resources, in the sense that a quantum circuit that outputs such states for arbitrary input states must contain an exponential number of elementary gates.
But exact randomness is not always necessary for all applications.  This leads to the study of $t$-designs,  where sets of states that can effectively simulate random states up to their $t$-th order statistical moments are investigated.
Realizing such sets in polynomial time has been studied in Refs.~\cite{EWSLC2003,AE2007,ODP2007, DOP2007, HL2009,PDP2008}.
In Refs.~\cite{ODP2007, DOP2007, PDP2008}, the average amount of entanglement is used as an indicator of the realizability of a $2$-design,
and it is shown that $2$-designs are implementable in polynomial time by random quantum circuits \cite{HL2009}, or
by a simple measurement procedure on a weighted graph state \cite{PDP2008}.

However, when we analyze random states for understanding the foundation of several models in physics \cite{GLTZ2006, PSW2006,LPSW2009,HP2007},
random quantum circuits and weighted graph states are rather artificial, since their implementations in physical systems require fine controls of many parameters in time. It is natural to ask if random states are approximately realizable
by a set of states randomly selected during time evolution governed by fundamental equations of motion such as the Schr$\ddot{\mathrm{o}}$dinger equation,
or a master equation.
This question has been addressed mainly in the field of quantum chaos
by investigating the time evolution of the amount of entanglement that is referred to as {\it entanglement evolution}.
It has been shown that the chaotic dynamics described by a time-dependent Hamiltonian can generate the same amount of entanglement as that of the typical entanglement on average \cite{SC2003, AV2006, B2009}.

Is time-dependent control of parameters necessary for simulating the typical entanglement of random states?  In this paper, to investigate this question, we consider a set of states randomly selected during the time evolution  generated by {\it time-independent} Hamiltonians and investigate the properties of the Hamiltonians for simulating the typical entanglement.  It is known that Hamiltonians composed of local interactions acting on a fixed number of consecutive particles cannot generate the same amount of entanglement as that of the typical entanglement in finite time from separable initial states \cite{HLW2006, BHV2006}.  For this reason, we have to consider many-body interactions.  In particular, we deal with one-dimensional spin-1/2 Ising-type Hamiltonians including various kinds of many-body interactions. We choose Ising-type Hamiltonians, since the simplicity of Ising-type models allows us to analytically investigate the enhancement of entanglement generation due to the many-body interactions. We note that recently, realizations of many-body interaction Hamiltonians have been studied, for example, using Rydberg atoms \cite{WMLZB2010}.

Since the time-independent Hamiltonian dynamics change only the phases of states in the eigenbasis, it cannot generate exact random states.   Our interest is how well the time evolution generated by the time-independent many-body interaction Hamiltonians approximates the properties of the typical entanglement of random states.
Due to the extremely high concentration of the typical entanglement, it is necessary to investigate Hamiltonians which can generate high entanglement on average.  The Ising-type Hamiltonians have the possibility to generate even larger amount of entanglement than that of the typical entanglement \cite{PhaseRandom}. This is another reason why we consider the Ising-type many-body interaction Hamiltonian.

In this paper, by examining the entanglement evolution for several kinds of many-body Ising-type interactions, we show that many-body interactions can dramatically enhance entanglement generation.  We also show that
the average and standard deviation of the entanglement distribution can be comparable to those of the typical entanglement in polynomial time for certain types of many-body interactions.

This paper is organized as follows. In Section \ref{Pre}, the measure of entanglement and the Hamiltonians investigated
in this paper are presented.   In Section \ref{hm}, we present the results of the entanglement evolution for Hamiltonians composed of neighboring $n$-body interactions.  In Section \ref{largehm}, we present the results of the entanglement evolution for Hamiltonians composed up to neighboring $n$-body interactions.  We compare the entanglement distributions of our models and the typical entanglement of random states in Section \ref{MR}.  We present a summary in Section \ref{Summary}.

\section{Preliminary} \label{Pre}

\subsection{Measure of entanglement}

The linear entropy of reduced density matrices is often used for analyzing the realizability of approximate random states, since it is a good indicator of the $2$-design \cite{ODP2007, DOP2007,PDP2008}.   Accordingly, we use the average linear entropy of one-site reduced density matrices,
known as the Meyer-Wallach measure of entanglement.  For a pure state $\ket{\Phi}$ of a $N$-spin system, the Mayer-Wallach measure is defined by
\begin{equation}
E_{\mathrm{MW}}(\ket{\Phi})=\frac{2}{N} \sum_{i=1}^N S_L(\rho_i), \notag
\end{equation}
where
$\rho_i:=\tr_{\neg i} \ketbra{\Phi}{\Phi}$ is the reduced density matrix at the $i$-th spin,
and $S_L(\rho):=1-\tr\rho^2$ is a linear entropy~ \cite{MW2002, B2003}.  (The partial trace $\tr_{\neg i}$ is taken for the degrees of freedom all spins except for the $i$-th spin.)

The Mayer-Wallach measure satisfies
$0 \leq E_{\mathrm{MW}}(\ket{\Phi}) \leq 1$.  $E_{\mathrm{MW}}(\ket{\Phi})=0$ if and only if
$\ket{\Phi}$ is a separable state, and
$E_{\mathrm{MW}}(\ket{\Phi})=1$ if and only if $\ket{\Phi}$ is local unitarily equivalent to a GHZ state.
For the typical entanglement of random states, the entanglement distribution according to the Meyer-Wallach measure is obtained in Ref. \cite{SC2003}.
Its average and standard deviation are given by
\begin{equation}
\Raverage{E_{\mathrm{MW}}}=1-\frac{3}{2^N+1}, \label{EntRandomStates}
\end{equation}
and
\begin{equation}
\Rsigma= O\left(\frac{1}{2^N}\right). \label{SDRandomStates}
\end{equation}

\subsection{Generalized Ising model} \label{gI}

We present generalizations of a one-dimensional spin-1/2 Ising Hamiltonian for many-body interactions.
We consider a one dimensional $N$-spin system, in which each spin is specified by the index of its site $i$,
and interactions of spins act only on consecutive spins.
We impose a periodic boundary condition. For simplicity, we set the Planck constant $\hbar$=1 in this paper.

First, we define a Hamiltonian with neighboring $n$-body interactions for $n \geq 2$ such as
\begin{equation} \label{hn}
h_n:=\sum_{j=1}^N J_j^{(n)} \sigma_j^Z \otimes \cdots \otimes \sigma_{j+n-1}^Z,
\end{equation}
where $\sigma_j^Z$ is a Pauli $Z$ operator acting on the spin at the $j$-th site, and
$J_j^{(n)}$ is a site-dependent coupling constant.
Next, we define the most general Ising-type
Hamiltonian $H_n$ for $n \geq 2$ composed of single-spin Hamiltonians and neighboring at most $n$-body interactions,
\begin{equation} \label{eqlargehn}
H_n:= H_1 + \sum_{m=2}^n \Delta_m h_m,
\end{equation}
where a constant $\Delta_m$ denotes the strength of each $h_m$ for $2 \leq m \leq n$ and $H_1:= \sum_{j=1}^N b_j \sigma_j^Z$ denotes the contributions of the single-spin Hamiltonians.
Since $H_1$ commutes with $h_m$ for all $m \geq 2$, it only generates additional local unitary operations and does not affect properties of entanglement. We set $H_1=0$ without loss of generality for investigating the entanglement evolution.

Note that in physics, it is natural to expect that many-body interactions are weaker than few-body interactions.
Thus, in many cases, we expect that a set of coefficients $\{ \Delta_m \}$ satisfies the relationship $1 > \Delta_m  >\Delta_{m'}$ for $1< m < m'$.  However, in special cases, this relationship is not necessary to be satisfied for all coefficients, since some of the coefficients have to vanish due to the symmetry of the system.  For instance, for parity invariant systems,
the odd terms $h_{2k+1}$ vanish.

The state at time $t$ is given by $\ket{\Psi(t)}=e^{-i H_n t} \ket{\Psi_0}$
where $\ket{\Psi_0}$ is an initial state.
We study the entanglement distribution of the states $\ket{\Psi(t)}$ by investigating
the infinite-time average of the amount of entanglement, defined by
\begin{equation}
\average{E_{\mathrm{MW}}}_{T;\infty}:=\lim_{T\rightarrow \infty} \frac{1}{T} \int_0^T \EMW(\ket{\Psi(t)}),\notag
\end{equation}
and the shortest time required for achieving $\average{E_{\mathrm{MW}}}_{T;\infty}$.

We choose a separable initial state $\ket{\Psi_0}$ that generates $\average{E_{\mathrm{MW}}}_{T;\infty}$
as high as possible.   In Appendix~\ref{InitialState},
under the assumption of phase ergodicity, where the phases of $\ket{\Psi(t)}$ are sufficiently randomized in the long-time limit,
it is shown that
\begin{equation}
\average{E_{\mathrm{MW}}}_{T;\infty} \leq 1-\frac{1}{2^{N-1}},\notag
\end{equation}
holds and the maximum value is achieved if and only if
the initial state $\ket{\Psi_0}$ is given by $\otimes_{i=1}^N \ket{+_i}$,
where $\ket{+_i}$ are the eigenstates of $\sigma^X_i$ with eigenvalue $+1$.
Since this phase ergodicity assumption does not necessarily hold for $H_n$ in general, the maximum value is not guaranteed, but the initial state gives at least a lower bound for the generated entanglement.
Thus, we fix the initial state to be $\ket{\Psi_0}=\otimes_{i=1}^N \ket{+_i}$.
In \cite{PhaseRandom}, it is shown that the maximum value of $\average{E_{\mathrm{MW}}}_{T;\infty}$ is greater than
the average amount of entanglement over the typical entanglement given by Eq.~\eqref{EntRandomStates}.
That is, there is a possibility for the states $\ket{\Psi(t)}$ to have an even
larger amount of entanglement on average than that of the typical entanglement.

We denote the entanglement distribution of a set of randomly selected states $\{ \ket{\Psi(t)} \}$ generated by the Hamiltonian dynamics by $\{ \EMW(t) \}_t$.   We compare it with that of the typical entanglement of random states
$\{ \EMW \}_{\mathrm{rand}}$.  We focus on the effects of
the order of the many-body interactions, $n$, and the distribution of coupling constants $\{J_i^{(m)} \}$ for $2 \leq m \leq n$.
In order to examine their effects independently,
we consider four Hamiltonians, $\bar{h}_n$, $h_n$, $\bar{H}_n$ and $H_n$,
where $\bar{h}_n$ and $\bar{H}_n$ are special cases of Hamiltonians of $h_n$ and $H_n$ with uniform coupling constants.
We denote the corresponding entanglement distributions by $\{ \EMW^{\bar{h}_n}(t) \}_t$, $\{ \EMW^{h_n}(t) \}_t$,
$\{ \EMW^{\bar{H}_n}(t) \}_t$ and $\{ \EMW^{H_n}(t) \}_t$.

\section{The case of neighboring $n$-body interactions} \label{hm}

In this section, we investigate the average and standard deviation of the entanglement distribution for the Hamiltonian composed of neighboring $n$-body interactions $h_n$ defined by Eq.~(\ref{hn}).  We also show the time evolution of the two-point correlation functions for $h_n$, which exhibit properties of correlations generated by the many-body interaction Hamiltonian. For simplicity, we drop the index $n$ of the coupling constant $J_j^{(n)}$ in Eq.~(\ref{hn}) and denote the coupling constants by $J_j$ in this section.

\subsection{Uniform coupling constants, $\bar{h}_n$ }\label{hm1}

We first investigate the case of the Hamiltonian $\bar{h}_n$, a special case of $h_n$ with uniform coupling constants, $J_i=J$. We denote the eigenstates of the Hamiltonian $\bar{h}_n$ by $\ket{a_1 \cdots a_N}:=\otimes_{i=1}^N \ket{a_i}$, where the index  $a_1 \cdots a_N$ is a sequence of binary numbers $a_i \in \{0,1 \}$ that correspond to the eigenvalues $\{-1, +1 \}$ of $\sigma^Z_i$.   The basis $\{ \ket{a_1 \cdots  a_N} \}$ is usually referred to as the computational basis. Then
$\bar{h}_n$ is written as
\begin{eqnarray}
\bar{h}_n &=& J \sum_{j=1}^N \sigma_j^Z \otimes \cdots \otimes \sigma_{j+n-1}^Z \nonumber \\
&=& J \sum_{\begin{subarray}{c}a_1,\cdots, a_N \\ =0,1 \end{subarray}} \sum_{j=1}^N
\prod_{k=j}^{j+n-1}(2a_k-1) \ketbra{a_1 \cdots a_N}{a_1  \cdots a_N}. \nonumber
\end{eqnarray}
From the initial state
\begin{equation} \label{initialstate1}
\ket{\Psi_0}=\otimes_{i=1}^N \ket{+_i}= 2^{-N/2} \sum_{\begin{subarray}{c}
a_1,\cdots, a_N \\ =0,1 \end{subarray}} \ket{a_1 \cdots a_N},
\end{equation}
the time evolution of the state $\ket{\Psi(t)}$ is described by
\begin{eqnarray}
\ket{\Psi(t)}=2^{-N/2} \sum_{\begin{subarray}{c}a_1,\cdots, a_N \\ =0,1 \end{subarray}} \prod_{j=1}^N
e^{-i J t \prod_{k=j}^{j+n-1}(2a_k-1)} \ket{a_1 \cdots a_N}. \nonumber
\end{eqnarray}

Due to the translational invariance of the Hamiltonian $\bar{h}_n$ and the initial state, all one-site reduced density matrices are identical. Thus, for any $l$-th spin, the reduced density matrix $\rho_l (t)=\tr_{\neg l} \ketbra{\Psi(t)}{\Psi(t)}$ is
\begin{equation}
\rho_l(t)= \frac{1}{2}
\begin{pmatrix}
1 & \cos^n 2Jt \\
\cos^n 2Jt & 1
\end{pmatrix}.\notag
\end{equation}
Then, we obtain
\begin{equation}
E_{\mathrm{MW}}(\ket{\Psi(t)})=1-\cos^{2n} 2Jt.\label{uniEMW}
\end{equation}
Note that $E_{\mathrm{MW}}(\ket{\Psi(t)})$ is independent of the system size $N$.

\begin{figure}[!tb]
   \centering
   \includegraphics[width=55mm, clip]{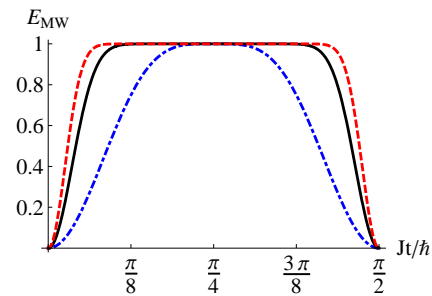}
   \caption{(Color  online) $E_{\mathrm{MW}}(\ket{\Psi(t)})$ generated by the uniform coupling constant Hamiltonian $\bar{h}_n$.  The blue dotted-dashed line, black solid line and red dashed line indicate $n=2,10,20$, respectively.
It takes exactly $t=\pi/4|J|$ to reach the maximum, independently of $n$.}
   \label{SingleUniform}
  \end{figure}

In this case, the entanglement evolution $E_{\mathrm{MW}}(\ket{\Psi(t)})$ for several different values of $n$ is shown in Fig.~\ref{SingleUniform}.   It is seen that the maximum $\EMW=1$ is reached only at $t=\pi/4|J|$ independently of $n$.
However, larger values of $n$ provide a faster speed to reach a neighborhood of the maximal value and a longer duration to stay in the the neighborhood of the maximal value.

As $E_{\mathrm{MW}}(\ket{\Psi(t)})$ is periodic with period $\pi/2|J|$,
it is easy to calculate the average amount of entanglement
\begin{align}
\average{E_{\mathrm{MW}}}_{T;\infty} &= \frac{1}{\pi/2|J|} \int_0^{\pi/2|J|} 1-\cos^{2n} 2Jt dt \notag\\
&= 1-\frac{(2n-1)!!}{(2n)!!} \notag\\
&\overset{n \rightarrow \infty}{\longrightarrow} 1-\frac{1}{\sqrt{n \pi}}.\notag
\end{align}
where $n!!=n(n-2)(n-4)\cdots$.
The standard deviation $\sigma_{T;\infty}$ is obtained by
\begin{align}
\sigma_{T;\infty} &=\sqrt{\frac{(4n-1)!!}{(4n)!!}-\left(\frac{(2n-1)!!}{(2n)!!}\right)^2} \notag\\
&\overset{n \rightarrow \infty}{\longrightarrow} \sqrt{\frac{1}{\sqrt{2n \pi}}-\frac{1}{n \pi}}. \notag
\end{align}
Because $\averageInf{\EMW}$ scales polynomially with $n$, the time evolution by $\bar{h}_n$ cannot
achieve the average amount of the typical entanglement even when $n$ is large.

\subsection{Site-dependent coupling constants, $h_n$} \label{hm2}

Next, we investigate the Hamiltonian $h_n$ defined by Eq.~(\ref{hn}), in which the coupling constants $J_j^{(n)}$ are site-dependent.  Since the eigenstates of $h_n$ are the same as those of $\bar{h}_n$,
the time evolution of the state from the initial state $\ket{\Phi_0}$ is given by
\begin{equation}
\ket{\Psi(t)}=2^{-N/2} \sum_{\begin{subarray}{c}a_1,\cdots, a_N \\ =0,1 \end{subarray}} \prod_{j=1}^N
e^{-i J_j t \prod_{k=j}^{j+n-1}(2a_k-1)} \ket{a_1 \cdots a_N}. \notag
\end{equation}
The reduced density matrix $\rho_l(t)$ of the $l$-th spin is
\begin{equation}
\rho_l(t)= \frac{1}{2}
\begin{pmatrix}
1 & A_l(t)\\
A_l(t) & 1
\end{pmatrix}, \notag
\end{equation}
where
\begin{equation}
A_l(t)=\prod_{k=l-n+1}^{l} \cos 2J_k t.
\end{equation}
Note that $\rho_l(t)$ now depends on $l$ since the Hamiltonian $h_n$ is no longer translationally invariant.
The entanglement evolution is
\begin{equation}
E_{\mathrm{MW}}(\ket{\Psi(t)})=1-\frac{1}{N} \sum_{l=1}^N (A_l(t))^2. \label{QQQ}
\end{equation}

Although $E_{\mathrm{MW}}(\ket{\Psi(t)})$ depends on the distribution of $\{ J_i \}$,
an upper bound of the long-time average is easily obtained by using
\begin{equation}
\lim_{T \rightarrow \infty} \frac{1}{T} \int_0^T \cos \theta t dt=
\begin{cases}
1 & (\text{if} \ \theta=0), \\
0 & (\text{if} \ \theta \neq 0).
\end{cases}\notag
\end{equation}
Then, the average amount of entanglement bounded by
\begin{align}
\average{E_{\mathrm{MW}}}_{T;\infty}&=
1-\frac{1}{N} \sum_{l=1}^N \averageInf{(A_l(t))^2} \nonumber \\
 &= 1- \frac{1}{2^n}\frac{1}{N} \sum_{l=1}^N \averageInf{\prod_{k=l-n+1}^l (1+\cos 4J_k t)} \nonumber \\
 &\leq1-\frac{1}{2^n}, \label{SingleSiteUpper}
\end{align}
where
$$\frac{1}{N}\sum_{l=1}^N \averageInf{\prod_{k=l-n+1}^l (1+\cos 4J_k t)} \geq 1$$
is used to evaluate the last expression.
For a completely random distribution of $\{ J_i \}$, the equality holds.

\begin{figure}
\centering
\begin{tabular}{c c}
  \includegraphics[width=40mm, clip]{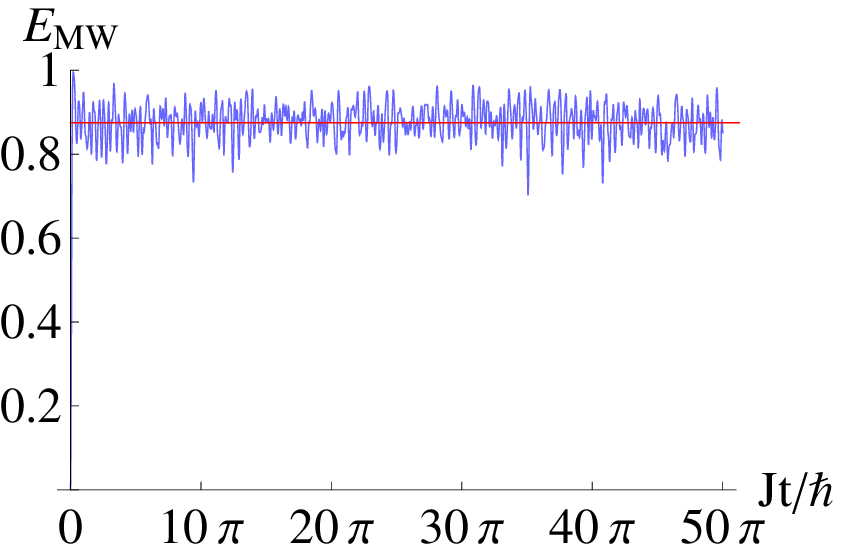} &   \includegraphics[width=40mm, clip]{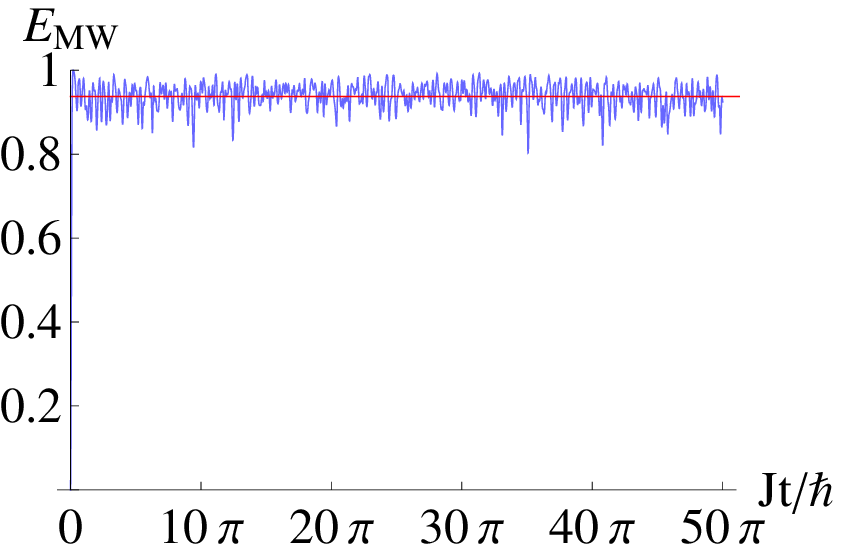}\\
  \footnotesize (a) & (b)\\
  \includegraphics[width=40mm, clip]{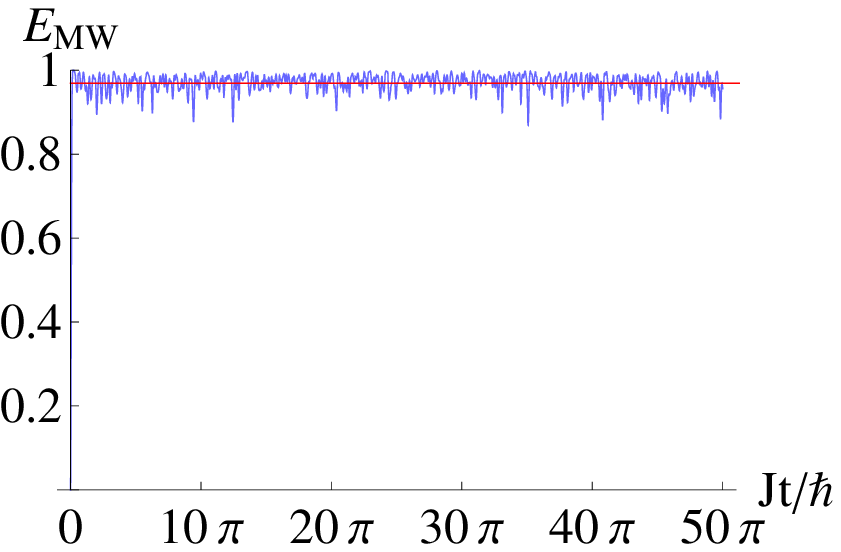} &   \includegraphics[width=40mm, clip]{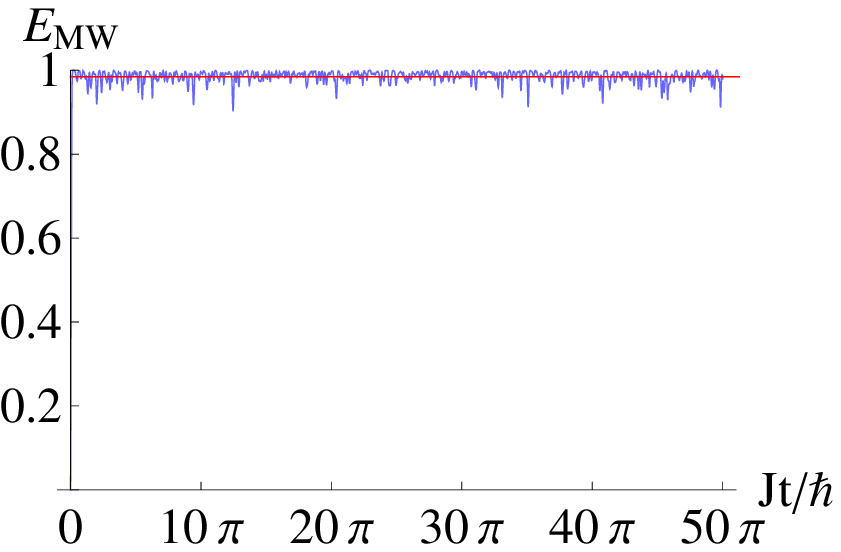}\\
  \footnotesize (c) & (d)
  \end{tabular}
  \caption{(Color  online)
$E_{\mathrm{MW}}(\ket{\Psi(t)})$ generated by site-dependent neighboring $n$-body interactions $h_n$ and for the number of spins $N=50$.
	The figures (a), (b), (c) and (d) correspond to $n=3, 4, 5, 6$, respectively.
	The red constant line in each figure represents the upper bound of the infinite time average given by Eq.~\eqref{SingleSiteUpper}.
	The distribution of $\{ J_i \}$ is set to Gaussian centered at $J$ with standard deviation $0.5 J$.}
	\label{SingleSiteDep}
\end{figure}

In Fig.~\ref{SingleSiteDep}, the entanglement evolution is shown when
the distribution of $\{ J_i \}$ is Gaussian centered at $J$ with standard deviation $0.5 J$.
It is seen that entanglement concentrates around the value given by Eq.~\eqref{SingleSiteUpper}.
The long-time average and its standard deviation are numerically obtained by
\begin{equation}  \label{EMWSingleSiteDep}
\average{\EMW}_{\tau_{\infty}} = 1-\frac{1.7}{2^{1.2n}},
\end{equation}
and
\begin{equation}
\sigma_{\tau_{\infty}} =  \frac{0.07}{2^{0.6 n}}. \label{SDSingleSiteDep}
\end{equation}
The average amount of entanglement $\average{\EMW}_{\tau_{\infty}}$ is greater than the upper bound given by
Eq.~\eqref{SingleSiteUpper} for large $n$, but this is merely an effect of finite time and it reaches the upper bound
in the infinite time average.

These values can be obtained in time of $O(1/J)$ that is deduced from Eq.~\eqref{QQQ}.
By comparing Eq.~\eqref{EMWSingleSiteDep} with the average amount of the typical entanglement given by Eq.~\eqref{EntRandomStates},
they coincide if $n \sim N$.
However, the standard deviation $\sigma_{\tau_{\infty}}$ does not coincide with
that of the typical entanglement given by Eq.~\eqref{SDRandomStates}.
We conclude that
a set of states obtained by the Hamiltonian dynamics with $h_n$
can simulate the typical entanglement in terms of
only the average amount of entanglement but not the standard deviation.
Comparing to the average entanglement generated by $\bar{h}_n$,
this result indicates that the randomness introduced in the large $n$-body interactions is necessary to generate high average entanglement.

\subsection{Two-point correlation functions}

In this subsection, we investigate the dynamics of correlation functions,
which are often discussed in association with entanglement in physics.
In order to clarify their connection, we examine two-point
spin correlation functions between the first spin and the $(r+1)$-th spin $C^{W}(r,t)$ for $W \in X, Y, Z$. Using a two-spin reduced density matrix of the first spin and the $(r+1)$-th spin denoted by $\rho_{1,r+1}$ and a single spin density matrix of the $i$-th spin denoted by $\rho_{i}$, $C^{W}(r,t)$ is written by
\begin{equation} \label{correlationfn}
C^{W}(r,t)=\tr \rho_{1,r+1} \sigma_1^W \sigma_{r+1}^W- (\tr \rho_1 \sigma_1^W)( \tr \rho_{r+1} \sigma_{r+1}^W),
\end{equation}
where the distance between the two spins $r$ is only taken for $r=1, \cdots, \frac{N}{2}-1$ due to the periodic boundary condition.

It is easy to see that for $C^{X}(r,t)$ and $C^{Y}(r,t)$, the first term of the right hand side of Eq.~\eqref{correlationfn} is a function of the elements $(\rho_{1,r+1})_{14}, (\rho_{1,r+1})_{23}, (\rho_{1,r+1})_{32}$ and $(\rho_{1,r+1})_{41}$ in the $\sigma^Z$ basis. On the other hand, the Meyer-Wallach measure is a function of the diagonal elements $(\rho_{1,r+1})_{ii}$ for $i=1,\cdots, 4$ and $(\rho_{1,r+1})_{13}$, $(\rho_{1,r+1})_{24}$, $(\rho_{1,r+1})_{31}$ and $(\rho_{1,r+1})_{42}$.
Thus,
$C^X(r,t)$ and $C^Y(r,t)$ reflect aspects of correlations of the system not included in the Meyer-Wallach measure.
On the other hand,
$\tr \rho_{1,r+1} \sigma_1^Z \sigma_{r+1}^Z$ is a function of the diagonal elements, so it is related to the Meyer-Wallach measure.

For the system evolving by the Hamiltonian $h_n$, the two-point spin correlation functions are analytically calculated to
\begin{align}
&C^{Z}(r,t)=0,  \notag \\
&C^{Y}(r,t)=0 \notag
\end{align}
and
\begin{align}
&C^{X}(r,t) \notag \\
&=
\begin{cases}
\displaystyle \prod_{l \in L} \cos2 J_l t -\prod_{\begin{subarray}{c} l\in P(N) \\ l'\in P(r) \end{subarray}}
\cos2 J_l t \cos2 J_{l'} t & (\text{if} \ r \leq n),  \\
0 & (\text{if} \ r >n),
\end{cases} \label{CorrX}
\end{align}
where
\begin{equation}
L=\begin{cases}
[N-n+2, N-n+r+1] \cup [2, r+1] \\
\ \ \ \ \ \ \ \ \ \ \ \ \ \  \ \ \ \ \ \ \ \ \ \  \ \ \ \ \ \ \ \ \ \  (\text{if}\ r< N-n+1), \\
[N-n+r+2, N-n+r+1] \cup [2, N-n+1] \\
\ \ \ \ \ \ \ \ \ \ \ \ \ \  \ \ \ \ \ \ \ \ \ \  \ \ \ \ \ \ \ \ \ \  (\text{if}\ r \geq N-n+1).
\end{cases}\notag
\end{equation}
and $P(k)=[k-n+2, k+1]$.  Except for the $X$ direction, distant spins cannot be correlated by the Ising-type Hamiltonian $h_n$.

\begin{figure}[tb]
\centering
  \includegraphics[width=65mm, clip]{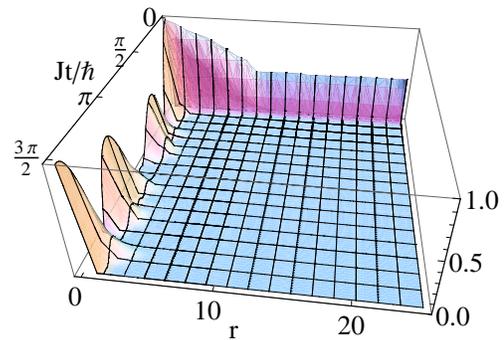}
  \caption{(Color  online) The correlation function $C^{X}(r,t)$ as a function of the distance between the spins $r$ and time $t$ for
the system evolving by the Hamiltonian $h_n$ and for $N=50$ and $n=40$.  When $t$ is small, two spins with large $r$ is correlated. However, the correlation for large $r$ rapidly decreases with time and disappears after a short time.}
  \label{CorrFig3D}
\end{figure}

\begin{figure}[tb]
\centering
\begin{tabular}{c c}
  \includegraphics[width=40mm, clip]{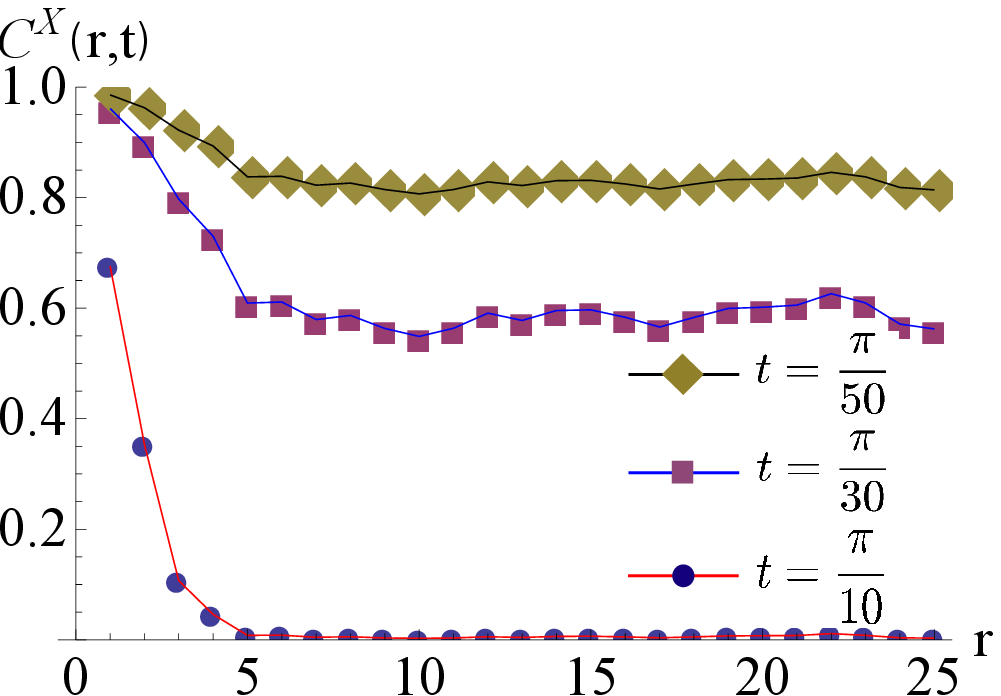} &   \includegraphics[width=40mm, clip]{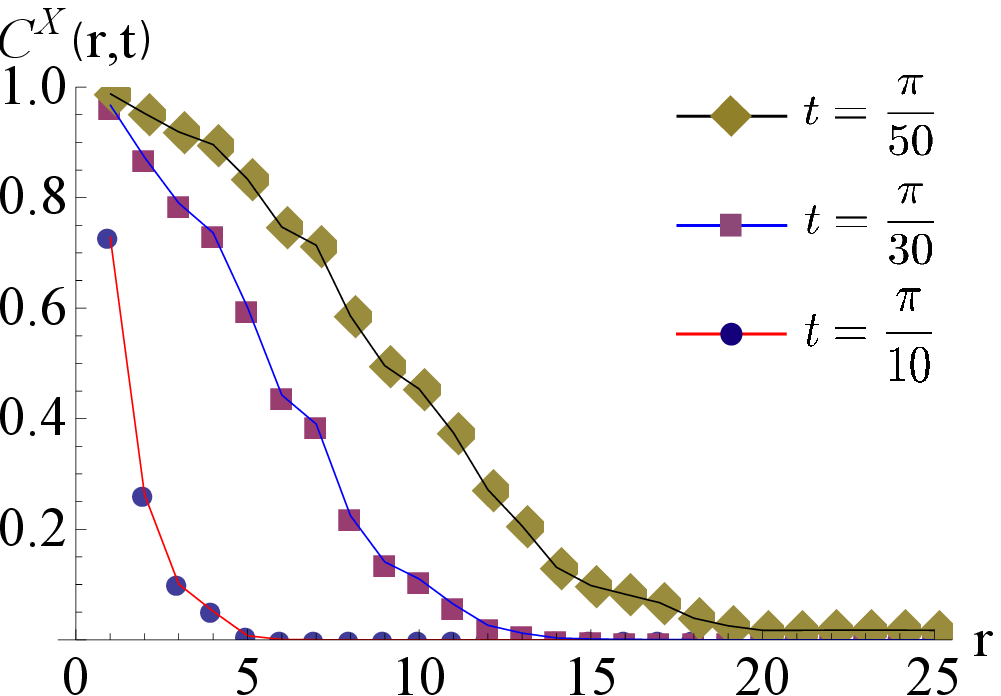}\\
  \footnotesize (a) & (b)\\
  \includegraphics[width=40mm, clip]{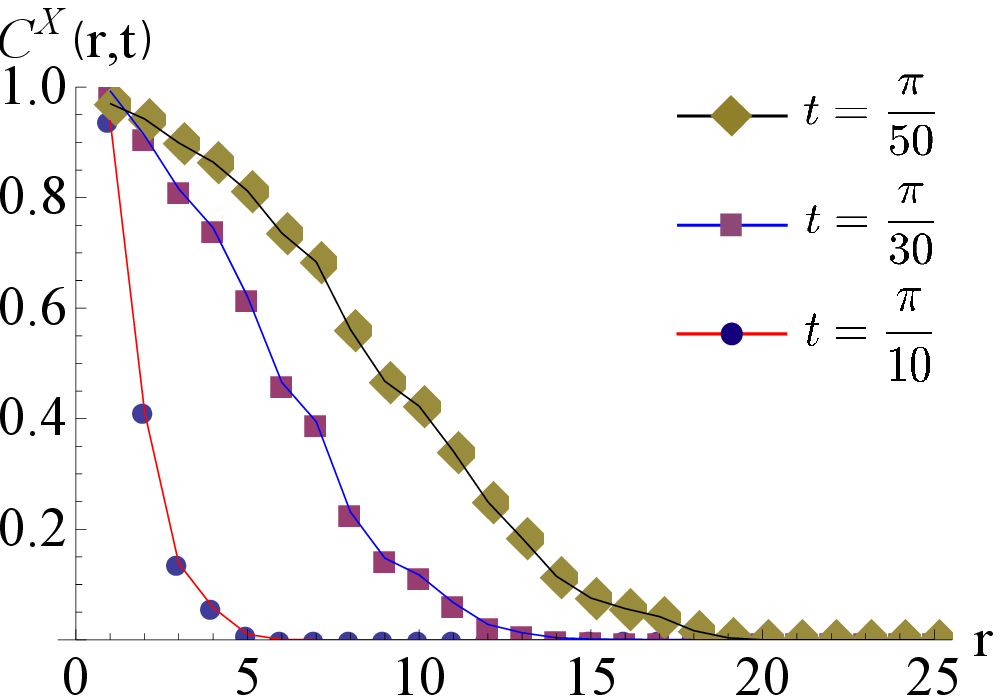} &   \includegraphics[width=40mm, clip]{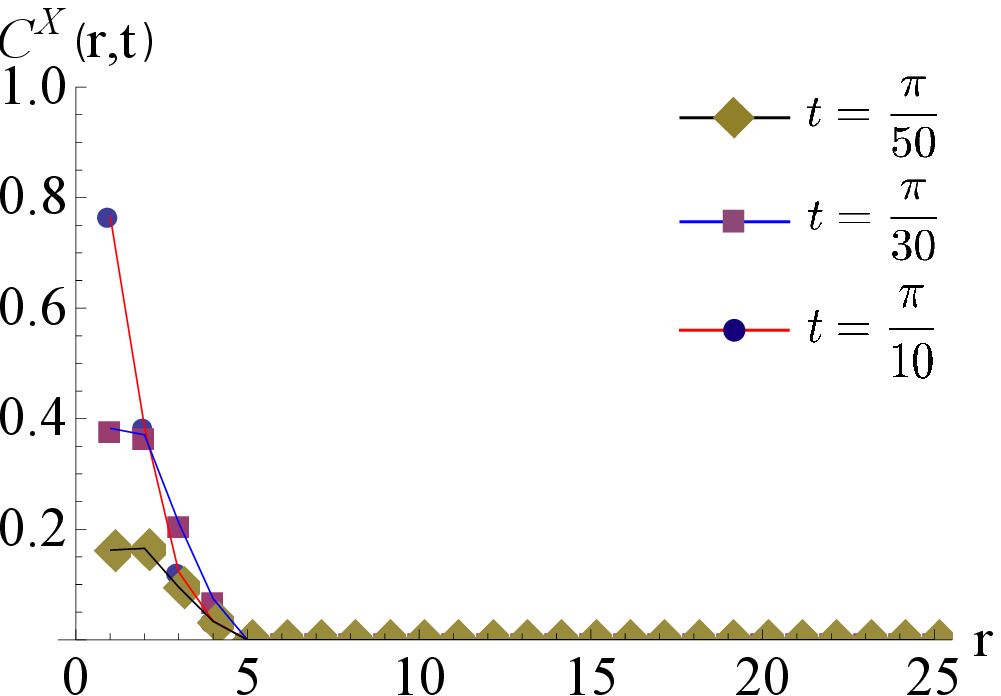}\\
  \footnotesize (c) & (d)
  \end{tabular}
  \caption{(Color  online) Correlation functions $C^{X}(r)$ as a function of the distance $r$ for a system of $N=50$ spins.
	The figures (a), (b), (c) and (d) show $n=45, 30,20, 5$, respectively.
	 In each figure, the diamond, square and circle points represent
  $t=\pi/(50J),\pi/(30J),\pi/(10J)$, respectively.
	It is seen that, for $r \geq n$, the correlation function is strictly zero in all cases, and that the correlation decreases quickly with time.}
	\label{CorrFig}
\end{figure}

In Fig.~\ref{CorrFig3D} and Fig.~\ref{CorrFig}, we show how $C^{X}(r,t)$ decreases with $r$ and $t$, depending on the order of interaction $n$.
When $n$ is small, $C^{X}(r,t)$ oscillates for small $r$ and is zero for large $r$,
which is implied by Eq.~\eqref{CorrX}.
On the other hand, as shown in Fig.~\ref{CorrFig3D}, when $n$ is large and $t$ is small, $C^{X}(r,t)$ is non-zero even for large $r$ .  Thus the large neighboring $n$-body interaction can create strong correlations between distant spins, as well as a large amount of entanglement.  However, in contrast to the entanglement evolution, the correlation rapidly decreases with time and the correlation between distant spins vanishes after a short time.  Thus, applying  the large neighboring $n$-body interaction does not dramatically enhance the two-point spin correlation functions in this model.

\section{The case of up to neighboring $n$-body interactions} \label{largehm}

In this section, we investigate the average and standard deviation of the entanglement distribution for the general Hamiltonian $H_n$,  composed of {\it up to} neighboring $n$-body interactions $h_n$, defined by Eq.~(\ref{eqlargehn}).

\subsection{Uniform coupling constants, $\bar{H}_n$ } \label{HnUni}

We first investigate the case of the Hamiltonian $\bar{H}_n$, a special case of  ${H}_n$ composed of up to  neighboring $n$-body interactions
with uniform coupling constants $J_j^{(m)}=J$.  The Hamiltonian is given by
\begin{equation}
\bar{H}_n=\sum_{m=2}^n \Delta_m \bar{h}_m.
\end{equation}
Since the Hamiltonian $\bar{H}_n$ is translationally invariant for uniform coupling constants,
the reduced density matrix $\rho_l (t)$ does not depend on $l$ and is in the form of
\begin{equation}
\rho(t)=  \frac{1}{2}
\begin{pmatrix}
1 & B(t) \\
B(t)^* & 1
\end{pmatrix},\notag
\end{equation}
where $B(t)$ is given by
\begin{multline}
B(t)=2^{1-N} \sum_{\begin{subarray}{c}a_1,\cdots, a_N \\ =0,1 \end{subarray}}
\prod_{m=2}^n \\
\times \prod_{j=n-m+1}^{n} \exp[-2 i \Delta_{m} \prod_{k=j}^{j+n-1}(2a_k-1) J t ], \notag
\end{multline}
and $B^*$ is the complex conjugate of $B$.
The entanglement evolution is
\begin{equation}
E_{\mathrm{MW}}(\ket{\Psi(t)})=1-|B(t)|^2, \notag
\end{equation}
which is independent of the number of spins $N$.

\begin{figure}
\centering
\begin{tabular}{c c}
  \includegraphics[width=40mm, clip]{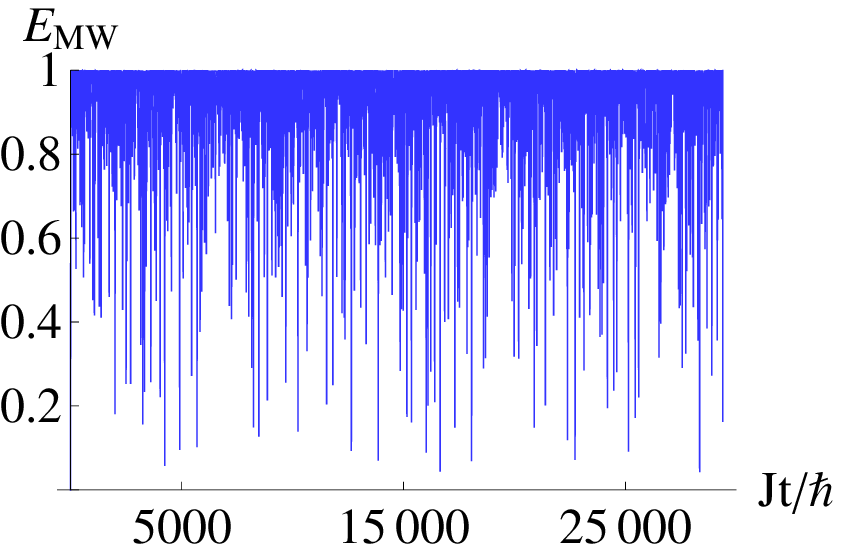} &   \includegraphics[width=40mm, clip]{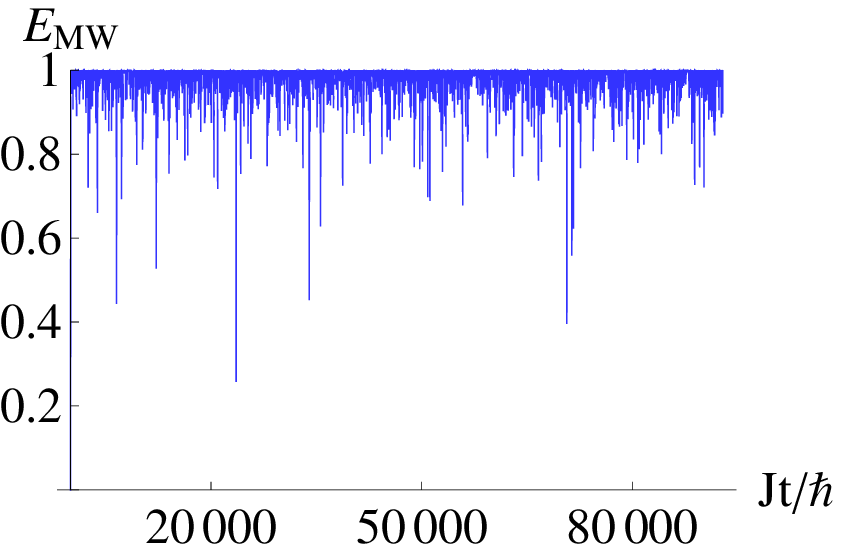}\\
  \footnotesize (a) & (b)\\
  \includegraphics[width=40mm, clip]{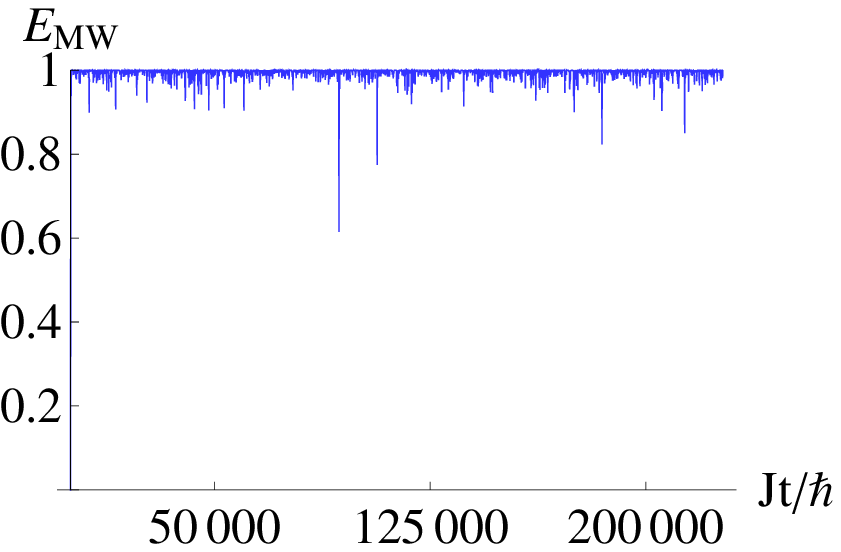} &   \includegraphics[width=40mm, clip]{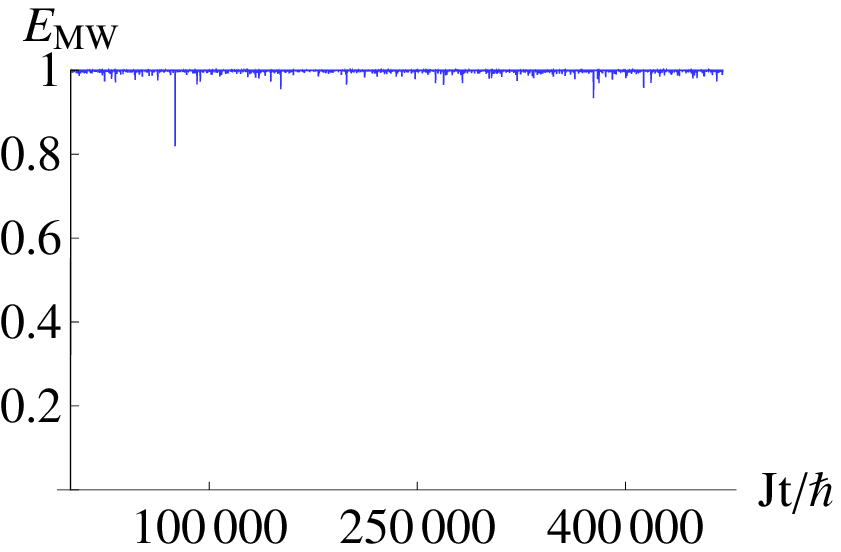}\\
  \footnotesize (c) & (d)
  \end{tabular}
  \caption{(Color  online) $E_{\mathrm{MW}}(\ket{\Psi(t)})$ generated by $\bar{H}_n$.  
	The figures (a), (b), (c) and (d) correspond to $n=3, 4, 5, 6$, respectively.
	The $\{ \Delta_m \}$ are set to $2 \epsilon/m$ with $\epsilon=\sqrt{3}/10$
	and the figures are plotted up to $t=(2 \pi n^4)/\epsilon$ which is $O(\mathrm{poly}(\Delta_n^{-1}))$.
	}\label{ManyUni}
\end{figure}

Although the average amount of entanglement is difficult to calculate in this case,
upper and lower bounds of the average amount of entanglement are analytically obtained by
\begin{equation}
1-\frac{4}{2^{\alpha (n-1)}} \leq \averageInf{\EMW} \leq 1-\frac{4}{2^{2 (n-1)}}, \label{ManyBound1}
\end{equation}
where $\alpha = \log_2 \frac{8}{3} \sim 1.42$ We show the derivation of these bounds in Appendix.~\ref{UpLow}.
Similarly, the standard deviation is estimated by
$$\sigma_{T;\infty}=O(2^{-2n}).$$
These values do not depend on the details of the strengths each $h_m$ and $\Delta_m$ in $\bar{H}_n$.
In Fig.~\ref{ManyUni}, the entanglement evolution generated by $\bar{H}_n$ with $\Delta_m=1/\mathrm{poly}(m)$
is plotted for $Jt \in[0, O(1/\mathrm{poly}(\Delta_n))]$, along with its time-average.  The entanglement evolution generated by the Hamiltonian with exponentially decreasing $\Delta_m$ behaves similarly to the case with polynomially decreasing $ \Delta_m $.

With numerical calculations, we obtain the average and standard deviation of the entanglement distribution
\begin{equation}
\average{\EMW}_{\tau_{\infty}} = 1-\frac{\bar{\beta}}{2^{\bar{\alpha} (n-1)}}, \label{EMWManyUni}
\end{equation}
and
\begin{equation}
\sigma_{\tau_{\infty}} = \frac{\bar{\gamma}}{2^{\bar{\delta} (n-1)}}\label{SDManyUni}
\end{equation}
where
\begin{equation}
(\bar{\alpha}, \bar{\beta}) =
\begin{cases}
 (2.0, 1.4) & (\text{if} \ \Delta_m =1/\mathrm{poly}(m)),\\
 (1.9, 1.6) & (\text{if} \ \Delta_m =1/\exp(m)),\label{constant}
\end{cases}
\end{equation}
and
\begin{equation}
(\bar{\delta}, \bar{\gamma}) =
\begin{cases}
 (1.8, 1.4) & (\text{if} \ \Delta_m =1/\mathrm{poly}(m)),\\
 (1.7, 5.5) & (\text{if} \ \Delta_m = 1/\exp(m)).
\end{cases}\notag
\end{equation}
These averages are taken over $0 \leq t \leq T_n$, where $T_n$ is chosen
to be a polynomial of $(J \Delta_n)^{-1}$.
Eq.~\eqref{constant} shows that the average amount of entanglement is not strongly dependent on $\{\Delta_m \}$,
which is expected from the bounds given by Eq.~\eqref{ManyBound1}.
The average $\average{\EMW}_{\tau_{\infty}}$ can be as high as that of the typical entanglement if
$n \sim N/\bar{\alpha}$, namely, $n \sim N/2 $.
In that case, the standard deviations are comparable with that of the typical entanglement as well.
Hence, if $n$ is set to $\sim N/2$,
the entanglement distribution obtained by the time evolution generated by $\bar{H}_n$ simulates
that of the typical entanglement
in terms of both the average and the standard deviation.

In order to investigate the time $\tau_{\infty}$ required to reach  $\averageInf{\EMW}$,
we consider the time evolution operator $\bar{U}_n=e^{-i \bar{H}_n t}$ which can be decomposed into
\begin{equation}
\bar{U}_n=e^{-i \Delta_2 \bar{h}_2 t}e^{-i \Delta_3 \bar{h}_3 t} \cdots e^{-i \Delta_n \bar{h}_n t}. \notag
\end{equation}
It is obvious that $e^{-i \Delta_n \bar{h}_n t}$ is almost the identity, if $\Delta_n t$ is
negligibly small compared to the largest eigenvalue of $\bar{h}_n$, which is equal to $NJ$.
Hence, it takes at least, $O( 1/(NJ \Delta_n))$ to achieve $\average{\EMW}_{T;\infty}$.
The numerical results shown in Fig.~\ref{ManyUni} support the claim that $O( 1/ (J \mathrm{poly}( \Delta_n)))$ is enough
to reach the infinite-time average.
For $n \sim N/2$, if $ \Delta_m $ scales polynomially with $m$, the time $\tau_{\infty}$ is a
polynomial in $N$.

\subsection{Site-dependent coupling constants, $H_n$}

\begin{figure}[!t]
\centering
\begin{tabular}{c c}
  \includegraphics[width=40mm, clip]{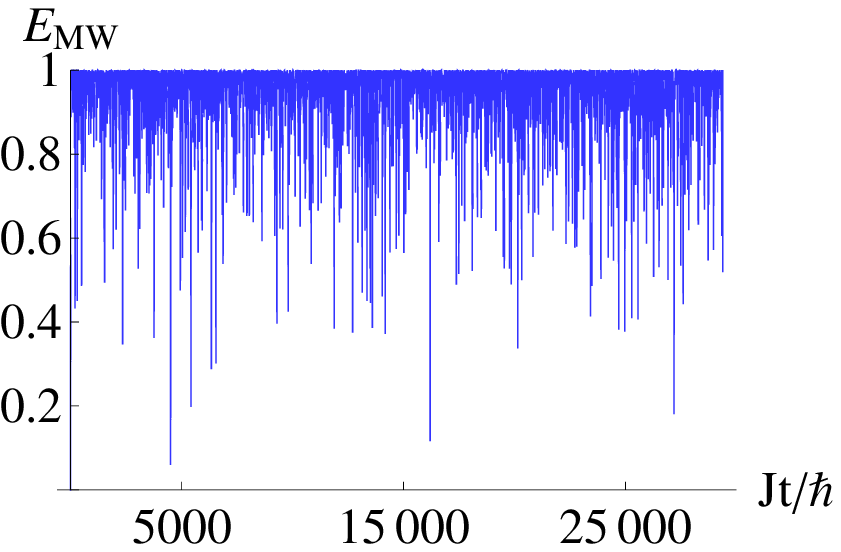} &   \includegraphics[width=40mm, clip]{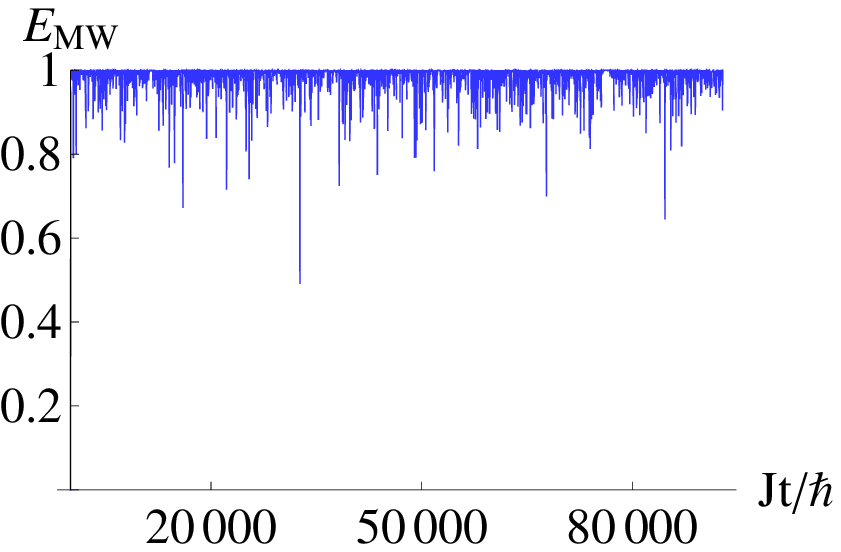}\\
  \footnotesize (a) & (b)\\
  \includegraphics[width=40mm, clip]{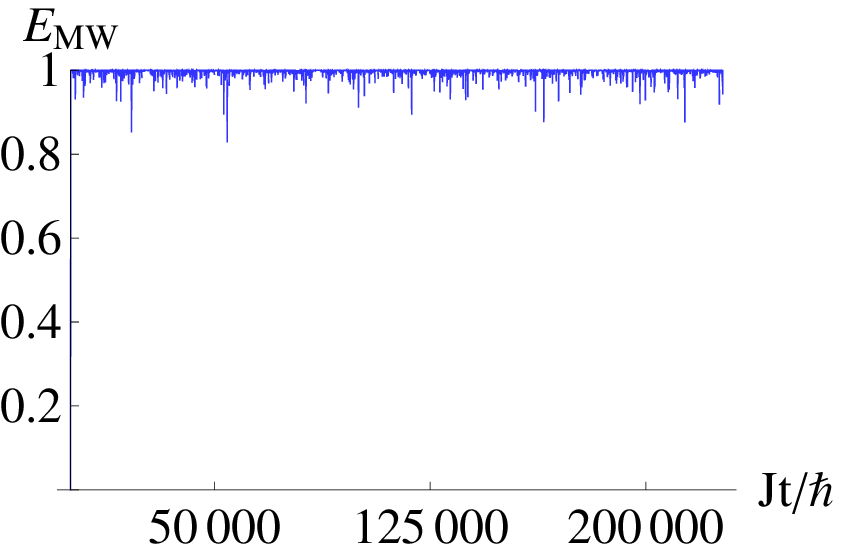} &   \includegraphics[width=40mm, clip]{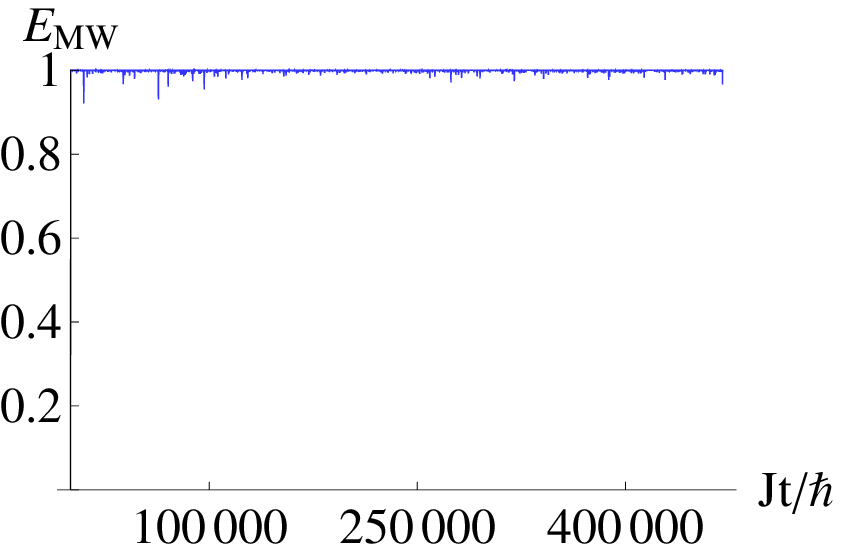}\\
  \footnotesize (c) & (d)
  \end{tabular}
  \caption{(Color  online) $E_{\mathrm{MW}}(\ket{\Psi(t)})$ generated by $H_n$. 
		The figures (a), (b), (c) and (d) correspond to $n=3, 4, 5, 6$, respectively.
	The plot is shown for the time scale $t \in [0,\mathrm{poly}(1/(J\Delta_n))]$.
	The distribution of $|J_i \}$ is given by a Gaussian centered at $J$ with standard deviation $0.5J$.
	The $\{ \Delta_m \}$ are set to $2 \epsilon/m $ with $\epsilon=\sqrt{3}/10$
	and the figures are plotted up to $t=(2 \pi n^4)/\epsilon$ which is $O(\mathrm{poly}(\Delta_n^{-1}))$.
	Comparing to Fig.~\ref{ManyUni}, the average amount of entanglement is higher.}
	\label{last}
\end{figure}

We show the result of the most general Hamiltonian $H_n$ defined by Eq.~(\ref{eqlargehn}).  In this case, each $h_m$ contained in $H_n$ has coupling constants depending on the site $i$, $J_i^{(m)}$.
$E_{\mathrm{MW}}(\ket{\Psi(t)})$ is obtained by
\begin{equation}
E_{\mathrm{MW}}(\ket{\Psi(t)})=1- \sum_{l=1}^N|B_l (t) |^2, \notag
\end{equation}
where $B_l (t)$ is given by
\begin{multline}
B_l (t) =2^{1-N} \sum_{\begin{subarray}{c}a_1,\cdots, a_N \\ =0,1 \end{subarray}}
\prod_{m=2}^n \\
\times \prod_{j=l-m+1}^{l} \exp[-2 i \Delta_{m-2} \prod_{k=j}^{j+n-1}(2a_k-1) J_l^{(m)} t ]. \notag
\end{multline}
The entanglement evolution as well as the average is shown in Fig.~\ref{last}.
Numerically, we obtain
\begin{eqnarray}
\average{\EMW}_{\tau_{\infty}} &= 1-\frac{1}{2^{2.12 (n-1)}}, \notag
\end{eqnarray}
and
\begin{eqnarray}
\sigma_{\tau_{\infty}} &= \frac{1.1}{2^{1.7 (n-1)}}.\notag
\end{eqnarray}
The strengths of the many-body interactions are chosen to be $\Delta_m = 2 \epsilon / m$ where $\epsilon=\sqrt{3} / 10$ and
the integral interval is taken from $0$ to $J t=2 \pi n^4 / \epsilon$ which is $O(1/(J \mathrm{poly}( \Delta_n)))$.

As is expected from the results of the previous subsections, both the average and the standard deviation of the entanglement distribution coincide to
those of the typical entanglement by choosing $n \sim N/2 $.
The time required for achieving a neighborhood of maximal entanglement is deduced from the same argument in Subsection \ref{HnUni},
and numerical calculations confirm that
$\tau_{\infty}=O(1/ (J \mathrm{poly}( \Delta_n)))$ is enough to achieve the infinite-time average
if $\{ \Delta_n \}$ scales polynomially with $n$.
That is, $\average{\EMW}_{T;\infty}$ and $\sigma_{T;\infty}$ are obtained in a time polynomial in $N$.

\section{Entanglement Distribution} \label{MR}

\begin{table}
\caption{
The time-averaged entanglement for each Hamiltonian defined in Section \ref{gI}.
$\averageInf{\EMW}$ is an approximation, found analytically, of the average in the infinite time limit.
Numerical results are given by $\average{\EMW}_{\tau_{\infty}}$ and $\sigma_{\tau_{\infty}}$,
which denote an average over the time interval $0$ to $\tau_{\infty}$ and its standard deviation, respectively.
The time required to reach the infinite-time average $\averageInf{\EMW}$ is denoted by $\tau_{\infty}$.} \label{TABLE}
{\linespread{2.5}\selectfont
\begin{tabular}{c||c|c|c|c}
 & $\bar{h}_n$ & $h_n$ & $\bar{H}_n$ & $H_n$ \\ \hline \hline

$\averageInf{\EMW}$ & $\displaystyle 1-\frac{1}{\sqrt{n \pi}}$ & $\displaystyle \sim 1-\frac{1}{2^n}$ &
$\displaystyle \sim 1-\frac{1}{2^{\alpha (n-1)}}$ &  - \\ 	

&&& ($\alpha \in [\log_2 \frac{8}{3}, 2] )$ & \\ \hline

$\displaystyle \average{\EMW}_{\tau_{\infty}}$ & - & $\displaystyle 1-\frac{1.7}{2^{1.2 n}}$ & $\displaystyle 1-\frac{1.4}{2^{2.0 (n-1)}}$ &  $\displaystyle 1-\frac{1.1}{2^{2.1 (n-1)}}$  \\ \hline

$\displaystyle \sigma_{\tau_{\infty}}$ & - & $\displaystyle \frac{0.07}{2^{0.6 n}}$ & $\displaystyle \frac{1.4}{2^{1.8 (n-1)}}$ &  $\displaystyle \frac{1.1}{2^{1.7 (n-1)}}$

\\ \hline

$\displaystyle \tau_{\infty}$ & $O(\frac{1}{J})$ & $O(\frac{1}{J})$ & $O(\frac{1}{N \Delta_n J})$& $O(\frac{1}{N \Delta_n J})$
\end{tabular} }
\end{table}

In this section, we compare the entanglement distributions obtained by Hamiltonians $\bar{h}_n$, $h_n$, $\bar{H}_n$ and $H_n$ and that of the typical entanglement of random states.   In Table.~\ref{TABLE}, we summarize the results obtained in the previous two sections.  First, we note that for all Hamiltonians we analyzed, the obtained entanglement
distributions depend on $n$ but not on $N$.

For the average of the entanglement distributions obtained by the Hamiltonians, it is shown that the Hamiltonians $h_n$, $\bar{H}_n$ and $H_n$ can generate entanglement as high as that of the typical entanglement by choosing suitable $n$.  For $h_n$, $n \sim N$ is required for the average to be comparable with $\Raverage{\EMW}$. For $\bar{H}_n$ and $H_n$, $n \sim N/2$ is sufficient for achieving $\Raverage{\EMW}$.  On the other hand, for the Hamiltonian composed of constant neighboring $n$-body interactions $\bar{h}_n$, the infinite-time average entanglement $\averageInf{\EMW}$ scales polynomially with $n$ even when $n$ is set to its maximum value $N$.

For the standard deviation of the entanglement distribution, the results for $\bar{h}_n$ and $h_n$ show that neighboring $n$-body interactions alone do not simulate the standard deviation of the typical entanglement. For $\bar{H}_n$ and $H_n$, $n \sim N/2$ is also sufficient for achieving the standard deviation coinciding to that of the typical entanglement of random states $\Rsigma$.  Thus these Hamiltonian dynamics can provide a set of states simulating the entanglement distribution of $\{\EMW \}_{\mathrm{rand}}$ in terms of up to the second order of distribution, namely, the average and standard deviation.

\begin{figure}[tb]
\centering
\begin{tabular}{c}
  \includegraphics[width=80mm, clip]{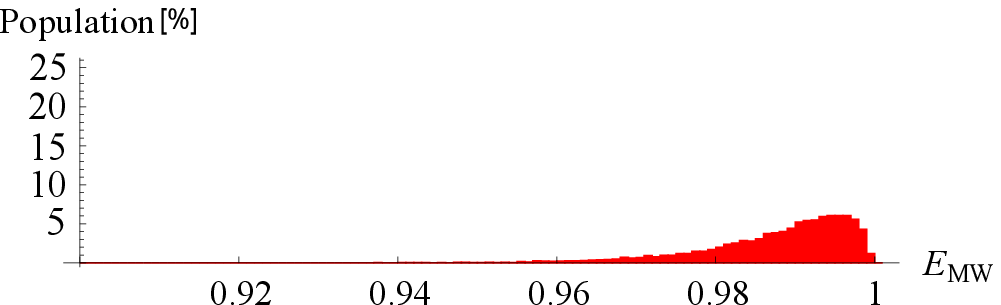}\\
  \footnotesize (a)\\
  \includegraphics[width=80mm, clip]{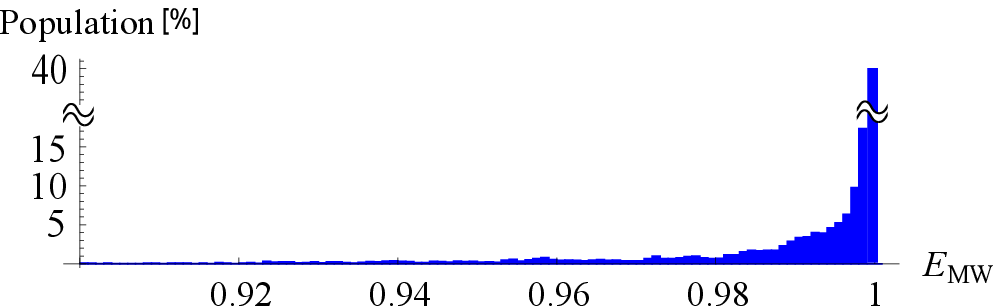}\\
  \footnotesize (b)\\
  \includegraphics[width=80mm, clip]{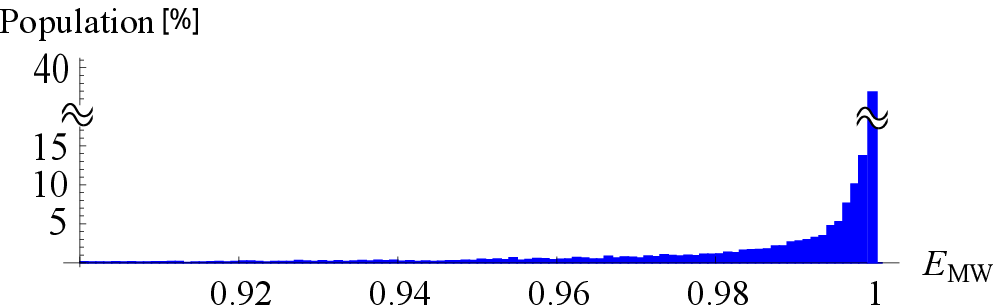}\\
  \footnotesize (c)
  \end{tabular}
  \caption{(Color  online) The entanglement distributions for $N=8$. The sampling number is 10000.
The figure (a) is the distribution obtained from the typical entanglement of random states, $\{ \EMW\}_{\mathrm{rand}}$.
The figure (b) is the distribution obtained by the Hamiltonian dynamics of $h_n$, $\{ \EMW^{h_n}(t) \}_t$, where $n=6$.
The figure (c) is the distribution obtained by the Hamiltonian dynamics of $H_n$, $\{ \EMW^{H_n}(t) \}_t$, where $n=4$.
Although the lower-order moments coincide with those of the typical entanglement, as shown in Table. \ref{TABLE},
		the entanglement distributions are different, which implies that the higher-order moments must be different.}
	\label{Distribution}
\end{figure}

However, these results do not guarantee that the entanglement distributions themselves obtained by the Hamiltonians coincide with that of the typical entanglement.  We show the calculations of the entanglement distributions for $N=8$ for $h_n$, $H_n$ and the typical entanglement in Fig.~\ref{Distribution}. For each Hamiltonian, $n$ is set to the value that achieves the average amount of the typical entanglement. The entanglement distributions obtained by these Hamiltonians are clearly different from that of the typical entanglement for $N=8$.  In these cases,  the entanglement distributions obtained by Hamiltonian dynamics is concentrated around the values higher than that of the distribution of the typical entanglement. Such properties of the entanglement distributions obtained by Hamiltonian dynamics may be useful for applications requiring to use a set of states with a higher concentration of entanglement in terms of Meyer-Wallach measure.

Next, we analyze the shortest time $\tau_{\infty}$ required to achieve  $\averageInf{\EMW}$.
For $\bar{h}_n$ and $h_n$, $\tau_{\infty}$ is $O(1/J)$. For $h_n$, $J$ is the center of Gaussian of the distribution of the coupling constant $\{ J_i \}$.  Thus the average amount of entanglement is achievable in a time scales independent of $N$.
On the other hand, for $\bar{H}_n$ and $H_n$, $\tau_{\infty}$ is inversely proportional to
the strength of the neighboring $n$-body interaction, $\Delta_n$, which is determined by the details of the Hamiltonian.
If $\Delta_m$ scales with $1/\mathrm{poly} (m)$, then $\tau_{\infty}=\mathrm{poly}(n)$.
Because $n \sim N/2$ is necessary in order to achieve $\Raverage{\EMW}$, the time required to achieve
$\averageInf{\EMW}$ is also polynomial in $N$.  On the other hand, if $\Delta_m$ scales with $1/\mathrm{exp} (m)$, the time required to achieve $\averageInf{\EMW}$ by the Hamiltonian dynamics is exponential in $N$.

\section{Summary} \label{Summary}

In this paper, we have studied the entanglement distributions obtained by one
dimensional spin-1/2 Ising-type Hamiltonians composed of several types of many-body interactions.
We have shown that, when the time-independent Hamiltonian is composed of up to $n \sim N/2$ neighboring $n$-body interactions, a set of states randomly selected during the time evolution generated by the Hamiltonian can simulate the entanglement distribution of
the typical entanglement of random states in terms of both the average and the standard deviation. 
Our results imply that $n \sim \log N$ neighboring $n$-body Ising-type interactions are not sufficient for generating high entanglement, even though the order of many-body interactions $n$ scales with the number of spins $N$ and the interactions are not local interactions in the sense defined by \cite{BHV2006}. We have also shown that the time required to achieve such a distribution is polynomial in the systems size if the strength of neighboring $m$-body interactions scales with $1/\mathrm{poly}(m)$.

On the other hand, when the Hamiltonian is composed of only neighboring $n$-body interactions,
site-dependent coupling constants and $ n \sim N $ neighboring $n$-body interactions are required to simulate the
average amount of entanglement of the typical entanglement. Although the standard deviation does not coincide to that of the distribution of the typical entanglement, the time required to achieve the distribution to simulate the average amount of entanglement alone is shown to be independent of the system size in this case.

\begin{acknowledgments}
The authors thank P. S. Turner for helpful discussions and A. Tanaka for useful comments.
This work was supported by Project for Developing Innovation Systems of
the Ministry of Education, Culture, Sports, Science and Technology (MEXT), Japan. Y.~N. acknowledges support from JSPS by KAKENHI (Grant No. 222812) and M.~M. acknowledges support from JSPS by KAKENHI  (Grant No. 23540463).
\end{acknowledgments}

\appendix

\section{Derivation of the initial state} \label{InitialState}

In this appendix, we show that the initial state $\ket{\Psi_0}=  \otimes_{i=1}^N \ket{+_i}$ gives the maximal average  amount of entanglement $\average{E_{\mathrm{MW}}}_{T;\infty}$, if the eigenstates of the Hamiltonian and the initial state are all separable.

In \cite{PhaseRandom}, we have introduced {\it phase-random states}, a set of  pure states $\{ \sum_{a=1}^{2^N}r_a e^{i \varphi_a} \ket{u_a} \}_{\{ \varphi_a \}}$ with fixed amplitudes $r_a \geq 0$ and uniformly distributed phases $\{ \varphi_a \}$ in a fixed basis $\{ \ket{u_a} \}$ and derive a formula for the average amount of linear entropy of entanglement of phase random states $\langle E_L \rangle_{\mathrm{phase}}$ for every possible bipartite partition.  
We have analyzed the entanglement evolution generated by a general Hamiltonian denoted by
$$H=\sum_{a=1}^{2^N} \varepsilon_a \ketbra{\varepsilon_a}{\varepsilon_a},$$
where $\{ \ket{\varepsilon_a} \}$ are eigenstates, from an initial state denoted by
\begin{eqnarray} \label{generalinitialstate}
\ket{\Psi(0)}= \sum_{a=1}^{2^N} r_a e^{ i  \omega_a} \ket{\varepsilon_a},
\end{eqnarray}
where $r_a, \omega_a \in \mathbb{R}$ and $\sum_a{| r_a |^2}=1$.  
Under the assumption of phase ergodicity in the sense that 
the distribution of phases of $e^{-i \varepsilon_a t+ i \omega_a}$ are uniform in $[0, 2\pi]$ in the long-time limit, the infinite-time average of the amount of entanglement $\langle E_L \rangle_{T; \infty}$ coincides to the average amount of entanglement of phase random states $\langle E_L \rangle_{\mathrm{phase}}$.   From the formula for $\langle E_L \rangle_{\mathrm{phase}}$, it is straightforward to derive the following lemma in terms of the Mayer-Wallach measure of entanglement.

\begin{Lemma}
If the system exhibits phase ergodicity, then
\begin{multline}
\average{E_{\mathrm{MW}}}_{T;\infty}= \average{E_{\mathrm{MW}}}_{\mathrm{phase}} \\
= \frac{2}{N}\sum_{k=1}^N[S_L(\rho^k_{\mathrm{av}}) + S_L(\rho^{\neg k}_{\mathrm{av}}) - S_L(\rho_{\mathrm{av}})] 
-\sum_{a=1}^{2^N} r_a^4 \EMW (\ket{\varepsilon_a}). \label{koredesu}
\end{multline}
where
$$\rho_{\mathrm{av}}=\sum_{a=1}^{2^N} r_a^2 \ketbra{\varepsilon_a}{\varepsilon_a},$$
$\rho_{\mathrm{av}}^{k}:=\mathrm{Tr}_{\neg k} \rho_{\mathrm{av}}$ and
$\rho_{\mathrm{av}}^{\neg k}:=\mathrm{Tr}_{k} \rho_{\mathrm{av}}$ for $k=1,\cdots, N.$
\end{Lemma}

We use this lemma for the Ising-type Hamiltonians and for separable initial states.  We denote the eigenstates of the Ising-type Hamiltonian $H_n$ in the Pauli $Z$ basis by using the binary representation $\ket{\bar{a}}$ defined by
\begin{eqnarray}  \label{binaryeigen}
\ket{\bar{a}} :=\ket{a_1 \cdots a_N}.
\end{eqnarray}
Using this notation, the time evolution of the state $\ket{\Psi (t)}$ generated by $H_n$ from a general initial state given by Eq.~(\ref{generalinitialstate}) is written by
\begin{equation}
\ket{\Psi(t)} =  \sum_{a=1}^{2^N} r_a e^{i (\omega_a - E_a t)} \ket{\bar{a}},
\end{equation}
where $E_a$ is the eigenenergy corresponding to $\ket{\bar{a}}$.

Assuming phase ergodicity, the average amount of entanglement $\average{E_{\mathrm{MW}}}_{T;\infty}$
is obtained by using the lemma.  Since the eigenstate $\ket{\bar{a}}$ is a separable state for all $\bar{a}$, $E_{\mathrm{MW}} (\ket{\bar{a}})=0$, $\average{E_{\mathrm{MW}}}_{T;\infty}$ is given by
\begin{equation}  \label{enw}
\average{E_{\mathrm{MW}}}_{T;\infty}=\frac{2}{N}\sum_{k=1}^N[S_L(\rho^k_{\mathrm{av}}) + S_L(\rho^{\neg k}_{\mathrm{av}}) - S_L(\rho_{\mathrm{av}})].
\end{equation}
where
$$\rho_{\mathrm{av}}=\sum_{a=1}^{2^N} r_a^2 \ketbra{\bar{a}}{\bar{a}}.$$

The right hand side of Eq.~(\ref{enw}) can be written in terms of $r_a$ by
\begin{multline}
\average{E_{\mathrm{MW}}}_{T;\infty}=\frac{4}{N} \sum_{k=1}^N [
\sum_{ \begin{subarray}{c} (i,m) \\ (j,n) \end{subarray}} r^2_{p_k(i,m)} r^2_{s_k(j,n)} \\
-\sum_{ (i,m)} r^2_{p_k(i,m)} r^2_{s_k(i,m)}  ], \label{EVALUATE}
\end{multline}
where $p_k(i,m):=i+2m \cdot 2^{N-k}$ and $s_k(i,m):=i+(2m+1) \cdot2^{N-k}$
and the summations are taken over $i,j=1, \cdots, 2^{N-k}$ and $m,n=0 ,\cdots 2^{k-1}-1$.
By using the normalization condition of $r_a$, the first term of Eq.~\eqref{EVALUATE} is evaluated as
\begin{align*}
&\frac{4}{N} \sum_{k=1}^N\sum_{ \begin{subarray}{c} (i,m) \\ (j,n) \end{subarray}} r^2_{p_k(i,m)} r^2_{s_k(j,n)}\\
&= \frac{4}{N} \sum_{k=1}^N \sum_{(i,m)} r^2_{p_k(i,m)}(1-\sum_{(j,n)} r^2_{p_k(j,n)})\\
&\leq 1.
\end{align*}
The equality holds if and only if $\sum_{(i,m)} r^2_{p_k(i,m)}=1/2$ for any $k$.

On the other hand, a lower bound of the second terms of Eq.~\eqref{EVALUATE} is given by
\begin{equation*}
\frac{4}{N} \sum_{k=1}^N \sum_{ (i,m)} r^2_{p_k(i,m)} r^2_{s_k(i,m)}
\geq 4 \left[ \prod_{k=1}^N \sum_{ (i,m)} r^2_{p_k(i,m)} r^2_{s_k(i,m)} \right]^{\frac{1}{N}},
\end{equation*}
since the arithmetic mean is always greater than or equal to the geometric mean.
The equality holds if and only if
$R(\mathbf{r}):=\sum_{ (i,m)} r^2_{p_k(i,m)} r^2_{s_k(i,m)}$ does not depend on $k$.
By substituting these relations,
the upper bound of the average is obtained by
\begin{equation*}
\average{E_{\mathrm{MW}}}_{T;\infty} \leq 1- 4 R(\mathbf{r}),
\end{equation*}
where the equality holds if and only if the two conditions,
\begin{align}
^{\forall} k, &\sum_{(i,m)} r^2_{p_k(i,m)}=1/2 \label{A} ,
\end{align}
and
\begin{align}
^{\forall} k, & \sum_{ (i,m)} r^2_{p_k(i,m)} r^2_{s_k(i,m)}=R(\mathbf{r}),  \label{B}
\end{align}
are simultaneously satisfied.

Since we consider a separable initial state $\ket{\Psi(0)}= \sum_{a=1}^{2^N} r_a e^{i \omega_a} \ket{\bar{a}}$,
it must satisfy the condition $\mathrm{rank} \tr_{\neg k} \ketbra{\Psi(0)}{\Psi(0)} =1$ for any $k$, that is,
\begin{equation}
^{\forall} k, ~
\mathrm{rank} \sum_{(i,m)}
\begin{pmatrix}
r^2_{p_k(i,m)} & \Omega_k (i,m) \\
\Omega_k^* (i,m) & r^2_{s_k(i,m)}
\end{pmatrix}
=1, \label{C}
\end{equation}
where $\Omega_k (i,m) := r_{p_k(i,m)} r_{s_k(i,m)} e^{i(\omega_{p_k(i,m)}-\omega_{s_k(i,m)})}$
and $\Omega_k^* (i,m)$ is the complex conjugate of $\Omega_k (i,m)$.
Under the conditions given by Eq.~\eqref{A} and Eq.~\eqref{C}, we minimize $R(\mathbf{r})$ so that
we can obtain the maximum value of the average $\average{E_{\mathrm{MW}}}_{T;\infty}$ for separable initial states.

From Eq.~\eqref{A} and Eq.~\eqref{C}, we obtain
\begin{equation}
|\sum_{(i,m)} \Omega_k (i,m)| =\frac{1}{2}. \notag
\end{equation}
With this condition, $R(\mathbf{r})=\sum_{(i,m)} |\Omega_k (i,m)|^2$ is easily maximized
by Lagrange's method of undetermined multipliers such as
\begin{equation}
R(\mathbf{r}) \geq \frac{1}{2^{N+1}}, \notag
\end{equation}
where the equality holds if and only if $|\Omega_k (i,m)|= 1/2^N$ and $\mathrm{Arg} \Omega_k (i,m)$ is constant
for any $k$, $i$ and $m$.
Therefore, we have proven that, for any separable initial states,
\begin{equation}
\average{E_{\mathrm{MW}}}_{T;\infty} \leq 1- \frac{1}{2^{N-1}}. \notag
\end{equation}
The equality holds if and only if $\sum_{(i,m)} r^2_{p_k(i,m)}=1/2$ for any $k$, and
$|\Omega_k (i,m)|= 1/2^N$ for any $k$, $i$ and $m$, which are
equivalent to $r_a=1/2^{N/2}$ and $\omega_a=\omega$ for any $a$.
Thus, the initial state $\ket{\Psi_0}=e^{i \omega} \otimes_{i=1}^N \ket{+_i} $ gives the
maximum amount of average of entanglement. By dropping the global phase $e^{i \omega}$, we obtain $\ket{\Psi_0}=\otimes_{i=1}^N \ket{+_i} $.

\section{Upper and lower bounds of the average entanglement for $\bar{H}_n$} \label{UpLow}

In this appendix, we show the following statement; for the Hamiltonian $\bar{H}_n$ defined by Eq.~(\ref{eqlargehn}),
\begin{equation}
1-\frac{4}{2^{\alpha (n-1)}} \leq \averageInf{\EMW} \leq 1-\frac{4}{2^{2 (n-1)}}, \notag
\end{equation}
where $\alpha = \log_2 \frac{8}{3} \sim 1.42$.

From the initial state $\ket{\Psi_0}$ given by Eq.~(\ref{initialstate1}), the time evolution of the state by the Hamiltonian $\bar{H}_n$ is formally written by
\begin{equation} \label{EnergyMW}
\ket{\Psi(t)} = \frac{1}{2^{N/2}} \sum_{a=1}^{2^N} e^{-i E_a t} \ket{\bar{a}},
\end{equation}
where $\ket{\bar{a}}$ is the binary representation of the eigenstates of $\bar{H}_n$ (and also $H_n$) introduced by Eq.~(\ref{binaryeigen}), and $E_a$ is the eigenenergy corresponding to $\ket{\bar{a}}$ given by
\begin{equation} \label{eigenenergy}
E_a=J\sum_{m=2}^n \Delta_m \sum_{j=1}^{N} \prod_{t=j}^{j+m-1} (2 a_t -1).
\end{equation}

Since the Hamiltonian $\bar{H}_n$ is translationally invariant, the Meyer-Wallach measure
 $E_{\mathrm{MW}}(\ket{\Psi(t)})$ is given by $E_{\mathrm{MW}}(\ket{\Psi(t)}) = 2 - 2 \tr \rho_1^2 (t)$. By calculating the reduced density matrix $\rho_1(t)$ of $\ket{\Psi(t)}$ given by Eq.~(\ref{EnergyMW}),  the Mayer-Wallach measure is obtained by
\begin{equation}
E_{\mathrm{MW}}(\ket{\Psi(t)})= 1- \frac{1}{2^{2(N-1)}} \sum_{a, b=1}^{2^{N-1}} \cos[(\epsilon_a-\epsilon_b) t], \label{epsilon}
\end{equation}
where $\epsilon_a=E_a-E_{a+2^{N-1}}$.  From Eq.~(\ref{eigenenergy}), $\epsilon_a$ is given by
\begin{equation}
\epsilon_a=J\sum_{m=2}^n \Delta_m \sum_{j=N-m+2}^{N+1} \prod_{t=j}^{N} (2 a_t -1) \prod_{s=2}^{j+m-1} (2 a_s -1). \label{yayakosi}
\end{equation}

To calculate the bounds of $\average{E_{\mathrm{MW}}}_{T;\infty}$, it is sufficient to evaluate the term $\average{\cos[(\epsilon_k-\epsilon_l) t]}_{T;\infty}$.
This term gives one for $\epsilon_k = \epsilon_l$  and zero for $\epsilon_k \neq \epsilon_l$.
By defining  a quantity $\Xi$ by
\begin{equation}
\Xi:=(\text{the number of $(k,l)$ such that $\epsilon_k = \epsilon_l$ }), \notag
\end{equation}
we obtain
\begin{equation}
\average{E_{\mathrm{MW}}}_{T;\infty}= 1- \frac{\Xi}{2^{2(N-1)}}. \notag
\end{equation}

From Eq.~(\ref{yayakosi}), we see that $\epsilon_a$ depends on $a_t$ for $t=N-n+2, N-n+3, \cdots, N-n+n$ where $n=1, \cdots N$, but it does not depend on $a_t$ for $t=1,n+1, \cdots, N-n+1$.  By introducing a vector notation $\vec{a}_{(N-n+2) \rightarrow n}:=(a_{N-n+2}, \cdots, a_{N},a_{2},\cdots, a_{n})$, $\epsilon_a- \epsilon_b$ in Eq.~\eqref{epsilon}  is only determined by $\vec{a}_{(N-n+2) \rightarrow n}$ and $\vec{b}_{(N-n+2) \rightarrow n}$.
We define a quantity $\xi$, which is the number of pairs  $(\vec{a}_{(N-n+2) \rightarrow n},
\vec{b}_{(N-n+2) \rightarrow n})$ satisfying $\epsilon_a = \epsilon_b$.  Then $\Xi$ can be expressed in terms of $\xi$ by
\begin{equation}
\Xi= \xi \times 2^{2(N-2n+2)}, \notag
\end{equation}
where the factor $2^{2(N-2n+2)}$ appears due to the choice of $a_t$ and $b_t$ for $t=1,n+1, \cdots, N-n+1$.
Recall that $\epsilon_a$ ($\epsilon_b$) is independent of $a_t$ ($b_t$) for $t=1,n+1, \cdots, N-n+1$.
Thus, we obtain
\begin{equation}
\average{E_{\mathrm{MW}}}_{T;\infty}= 1- \frac{\xi}{2^{4n-6}}.\notag
\end{equation}

We evaluate the upper and lower bounds of $\xi$.  For the lower bound, when $a=b$, $\epsilon_a=\epsilon_b$ is trivially satisfied.
Since the number of choices of $\vec{a}_{(N-n+2) \rightarrow n}$ is $2^{2(n-1)}$,
we obtain $2^{2(n-1)} \leq \xi$. It is cumbersome to derive the upper bound of $\xi$, but
direct investigation of the expression of Eq.~\eqref{yayakosi} provides a bound $\xi \leq 6^{n-1}$.
Therefore, the upper and lower bounds of $\average{E_{\mathrm{MW}}}_{T;\infty}$ are
\begin{equation}
1-4 \left(\frac{3}{8} \right)^{n-1} \leq \average{E_{\mathrm{MW}}}_{T;\infty} \leq 1-4 \left(\frac{1}{4} \right)^{n-1}. \notag
\end{equation}

\end{document}